\documentclass[%
 reprint,
 amsmath,amssymb,
 aps,
]{revtex4-2}

\usepackage{graphicx}
\usepackage{dcolumn}
\usepackage{bm}
\usepackage{booktabs}

\begin{document}

\preprint{APS/123-QED}

\title{Broken Symmetry-driven Weyl Semimetal Phase in Zn-Substituted EuMn$_2$Sb$_2$}

\author{Deep Sagar$^{1}$}
\author{Arti Kashyap$^{1}$}
\email{arti@iitmandi.ac.in}
\affiliation{$^{1}$School of Physical Sciences, Indian Institute of Technology Mandi, Mandi--175005, India}




\date{\today}

\begin{abstract}
The interplay between magnetism and electronic topology offers a powerful route to realizing emergent quantum phases. Here, we show that Zn substitution in the layered compound EuMn$_2$Sb$_2$ drives a transition from a C-type antiferromagnetic semiconductor to an intrinsic magnetic Weyl semimetal. Using first-principles calculations, we demonstrate that the parent compound hosts a gapped antiferromagnetic ground state, while Zn substitution alters the magnetic exchange interactions and stabilizes ferromagnetism. In the spin--orbit-coupled regime, the coexistence of broken time-reversal ($\mathcal{T}$) and inversion ($\mathcal{P}$) symmetries leads to the formation of Weyl nodes near the Fermi level. These nodes act as monopoles of Berry curvature and give rise to topologically protected Fermi-arc surface states. Our results identify EuMnZnSb$_2$ as a tunable platform where magnetism and topology are intrinsically coupled, and establish chemical substitution as a viable strategy to engineer magnetic Weyl semimetals in correlated electron systems, with potential implications for spintronic and topological transport phenomena.

\end{abstract}

\maketitle


\section{\label{sec:level1}INTRODUCTION}

The interplay between magnetism and electronic topology, as well as phenomena such as spontaneous symmetry breaking, phase transitions, and the reconstruction of electronic structure, has emerged as a central theme in condensed matter physics.~\cite{hasan2010colloquium,qi2011topological,armitage2018weyl}. These fundamental processes not only deepen our understanding of emergent quantum states but also underpin the development of advanced magnetic and topological materials with potential applications in spintronics and quantum technologies~\cite{tokura2019magnetic,pesin2012spintronics}. In particular, magnetic topological semimetals have attracted considerable attention, as they possess the capacity to host exotic quasiparticles such as Weyl fermions.~\cite{armitage2018weyl,burkov2016topological}. These systems are characterized by non-degenerate band-crossing points in momentum space, where the conduction and valence bands intersect linearly, giving rise to unusual transport phenomena—such as the anomalous Hall effect and topologically protected surface states~\cite{wan2011topological,xu2015discovery}. In three-dimensional (3D) systems, including Weyl and Dirac semimetals, such band crossings are stabilized by symmetry considerations~\cite{young2012dirac,wang2012dirac}, where the breaking of either time-reversal symmetry ($\mathcal{T}$) or inversion symmetry ($\mathcal{P}$) is a necessary condition for the emergence of Weyl nodes~\cite{Murakami2007phase,wan2011topological,zhang2013topology}. Magnetic materials, in particular, provide a natural platform for realizing $\mathcal{T}$-symmetry breaking, and when combined with strong spin--orbit coupling (SOC) or $\mathcal{P}$ symmetry breaking, they can host intrinsic magnetic Weyl semimetal (MWSM) phases~\cite{chang2018magnetic,burkov2011weyl}.

Layered Mn-based 122 pnictides, $\mathrm{EuMn_2Pn_2}$ ($\mathrm{Pn = As, P, Sb, Bi}$), crystallizing in the trigonal CaAl$_2$Si$_2$-type structure (space group $P\bar{3}m1$), represent a promising class of materials for exploring the coupling between magnetism and electronic topology~\cite{ovchinnikov2025pnictides,ruhl1979new,anand2016metallic}. The parent compound $\mathrm{EuMn_2Sb_2}$ adopts this layered structure and exhibits competing magnetic interactions, where Mn moments order at high temperatures, while Eu moments undergo antiferromagnetic ordering at low temperatures, reflecting the coexistence of multiple magnetic sublattices~\cite{schellenberg2010121sb}. These compounds host localized Eu-$4f$ moments coexisting with Mn-$3d$ electrons, resulting in complex magnetic interactions and diverse electronic phases~\cite{dahal2019spin}. For example, EuMn$_2$As$_2$ exhibits an insulating ground state with multiple antiferromagnetic transitions associated with the Mn and Eu sublattices, highlighting the coexistence of localized and itinerant magnetism~\cite{anand2016metallic}. Similarly, EuMn$_2$P$_2$ and CaMn$_2$Sb$_2$ are reported as antiferromagnetic insulators with strong Mn-driven magnetic ordering and proximity to electronic delocalization transitions, indicating significant correlation effects~\cite{berry2023bonding,simonson2012magnetic}. In contrast, related compounds such as EuZn$_2$Sb$_2$ and EuZn$_2$As$_2$ exhibit semiconducting behavior with Eu-driven magnetism, demonstrating that the transition-metal site plays a crucial role in tuning both magnetic and electronic properties~\cite{wang2017single,weber2006low,zhang2008new,may2012properties,wang2022anisotropy}.

Recent studies have shown that these systems can host nontrivial topological phases when magnetism and SOC are appropriately tuned. In particular, EuCd$_2$Sb$_2$ has been identified as a prototypical system where magnetic exchange interactions drive a transition from an antiferromagnetic state to a Weyl semimetal (WSM) phase, accompanied by large Berry curvature and Fermi-arc surface states~\cite{ma2019spin,su2020magnetic}. Similarly, theoretical investigations on EuMn$_2$Bi$_2$ indicate that small variations in magnetic ordering or chemical composition can induce transitions between trivial insulating, Dirac, and WSMS phases, underscoring the strong sensitivity of topology to magnetic structure~\cite{choudhury2024emerging}. In our recent work on EuMnXBi$_2$ (X = Mn, Fe, Co, Zn), we demonstrated that chemical substitution can effectively tune magnetic ground states and induce topological phase transitions, including the emergence of WSMS phases driven by SOC and magnetic exchange interactions~\cite{sagar2026tunable}. More generally, WSMS phases have been realized in both nonmagnetic and magnetic systems through symmetry breaking~\cite{armitage2018weyl}. For instance, Weyl nodess can emerge at topological phase transition points in systems such as TlBiSe$_2$-based alloys via $\mathcal{P}$ symmetry breaking~\cite{Murakami2007phase,singh2012topological}, while noncentrosymmetric compounds such as SrSi$_2$ host Weyl nodess due to intrinsic structural chirality and SOC~\cite{huang2016new,sadhukhan2021electronic}. In magnetic systems, materials such as Co$_3$Sn$_2$S$_2$ have been established as prototypical MWSMs, exhibiting large anomalous Hall conductivity arising from Berry curvature near Weyl nodess close to the Fermi level~\cite{liu2018giant,morali2019fermi,xu2018topological,lohani2023electronic}. Furthermore, theoretical predictions in the RAlGe family demonstrate that the simultaneous breaking of $\mathcal{T}$ and $\mathcal{P}$ symmetries can lead to robust and tunable Weyl phases, where ferromagnetism effectively shifts Weyl nodess in momentum space~\cite{chang2018magnetic}.

Despite these advances, the realization of tunable MWSMs in Mn-based layered pnictides remains largely unexplored. In particular, the role of chemical substitution in driving simultaneous magnetic and topological phase transitions, especially those involving both $\mathcal{T}$ and $\mathcal{P}$ symmetry breaking, is not yet fully understood. Addressing this gap is essential for establishing design principles for magnetic topological materials based on correlated electron systems.

In this work, we employ density functional theory (DFT) calculations to investigate the magnetic and electronic properties of pristine $\mathrm{EuMn_2Sb_2}$ and the Zn-substituted compound $\mathrm{EuMnZnSb_2}$. We show that pristine $\mathrm{EuMn_2Sb_2}$ is a semiconducting system with a stable C-type antiferromagnetic ground state. Upon Zn substitution, the magnetic exchange interactions are significantly modified, leading to a transition from antiferromagnetic to ferromagnetic ordering and a concomitant reconstruction of the electronic band structure. This transition provides a pathway for realizing a MWSMs phase driven by substitution-induced symmetry breaking, establishing $\mathrm{EuMnZnSb_2}$ as a promising platform for exploring tunable magnetic topological states.

\begin{figure*}
    \centering
    \includegraphics[width=1\linewidth]{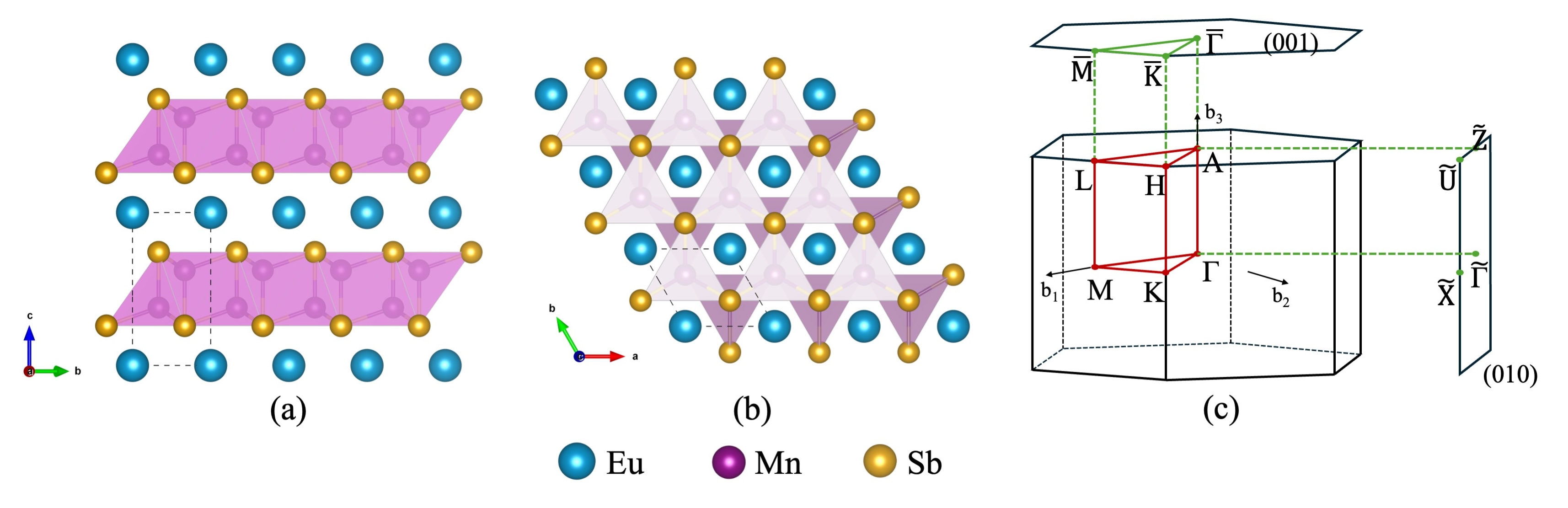}
    \caption{Crystal structure of EuMn$_2$Sb$_2$. (a) Side view and (b) top view of the conventional unit cell, highlighting the layered arrangement of Mn–Sb polyhedra separated by Eu layers. The dashed lines indicate the primitive unit cell. (c) Three-dimensional first Brillouin Zone (BZ) with the high-symmetry path marked in red, together with the projected two-dimensional BZs for the (001) and (010) planes. Blue, purple, and yellow spheres denote Eu, Mn, and Sb atoms, respectively.}
    \label{fig:placeholder}
\end{figure*}

\begin{figure}
    \centering
    \includegraphics[width=1\linewidth]{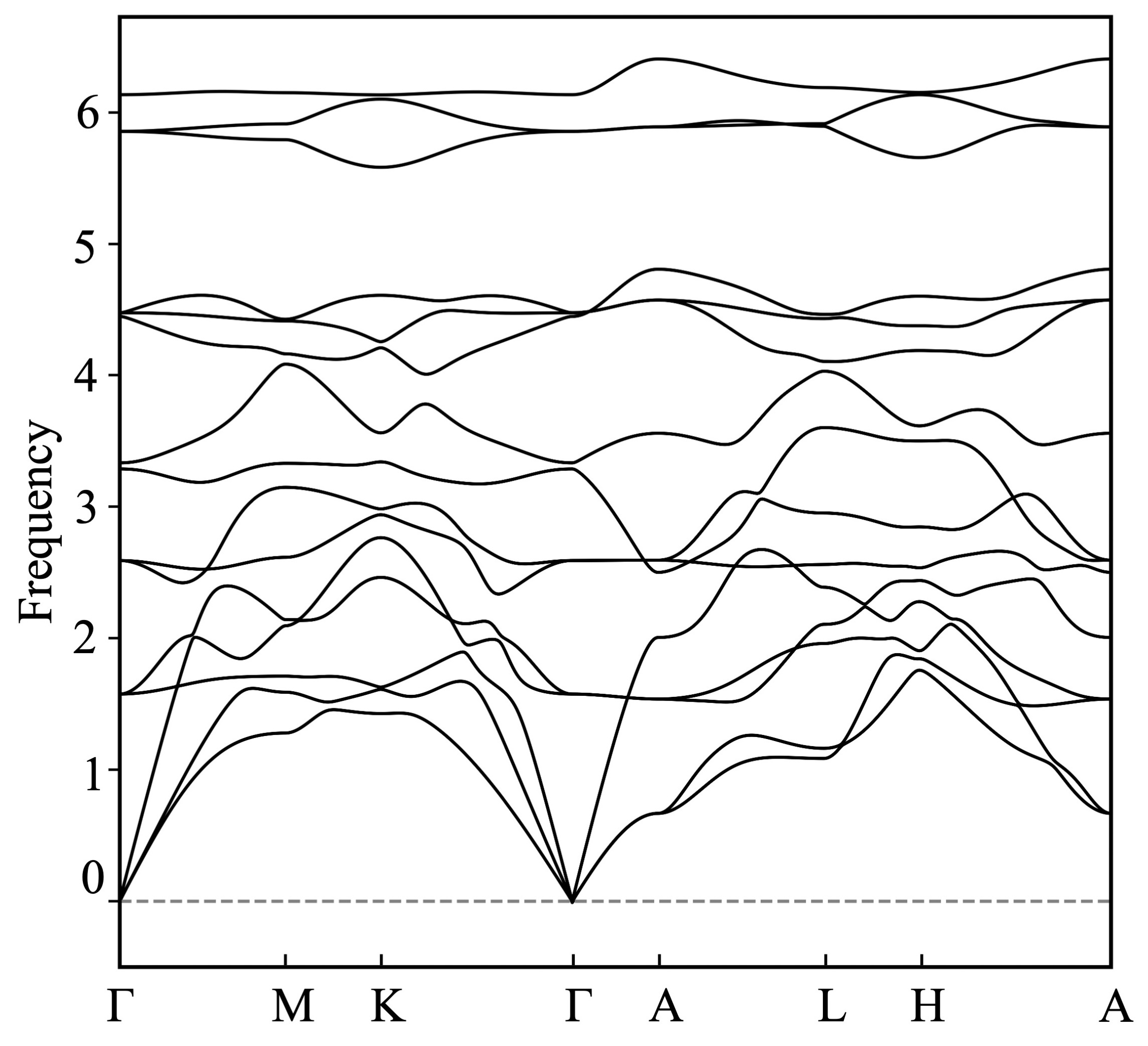}
    \caption{Phonon band structure. }
    \label{fig:placeholder}
\end{figure}

\section{\label{sec:level2}Computational Methods}

All calculations were performed within the framework of Density Functional Theory, using the Projector Augmented-Wave method, as implemented in the Vienna \textit{Ab initio} Simulation Package (VASP)~\cite{kresse1993ab,kresse1996efficient}. Exchange-correlation effects were analyzed under the generalized gradient approximation using the Perdew-Burke-Ernzerhof functional~\cite{perdew1996generalized}. The crystal structures were optimized by sampling the Brillouin zone with a $5 \times 5 \times 3$ $k$-point mesh~\cite{monkhorst1976special}. For improved accuracy in the electronic structure calculations, a denser $10 \times 10 \times 6$ $k$-point grid was employed. A plane-wave kinetic energy cutoff of 520~eV was used throughout. Electronic occupations were handled using Gaussian smearing with a width of 0.05~eV. Structural relaxation was performed until the forces acting on each atom were less than $1 \times 10^{-3}\,\mathrm{eV/\AA}$, and the total energy convergence criterion was set to $10^{-6}$~eV. Phonon dispersions and lattice dynamical properties were evaluated using the PHONOPY package~\cite{togo2015first} within the finite-displacement approach based on the Parlinski--Li--Kawazoe method~\cite{parlinski1997first}. To account for the localized nature of Eu $4f$ and Mn $3d$ electrons, on-site Coulomb interactions were included within the DFT+$U$ scheme~\cite{dudarev1998electron}. These corrections were found to significantly influence the electronic structure and magnetic properties of EuMn$_2$Sb$_2$ and EuMnZnSb$_2$. Spin--orbit coupling was incorporated self-consistently to capture relativistic effects. To investigate the topological characteristics, maximally localized Wannier functions were constructed using Wannier90~\cite{mostofi2008wannier90,marzari1997maximally}, from which an effective tight-binding Hamiltonian was derived. This Hamiltonian was further analyzed using WannierTools~\cite{wu2018wanniertools} to compute Berry curvature distributions, identify Weyl nodes, and obtain surface spectral functions, including Fermi arc states.

\section{\label{sec:level2}Results and Discussion }

\subsection{\label{sec:citeref}Structural Optimization}
The structural parameters for EuMn$_2$Sb$_2$ were taken from Refs.~\cite{ruhl1979new, schellenberg2010121sb}. EuMn$_2$Sb$_2$ crystallizes in a trigonal CaAl$_2$Si$_2$-type structure, with the space group P$\bar{3}$m1 (No. 164); it belongs to the hexagonal primitive (hP) Bravais lattice and point group $D{3d}$. Its optimized lattice parameters are $a = b = 4.491$ Å and $c = 7.657$ Å, and the lattice angles are $\alpha = \beta = 90^\circ$ and $\gamma = 120^\circ$; this yields a unit-cell volume of $133.74$ Å$^{3}$, which is consistent with previous experimental results of $a = b = 4.581$ Å and $c = 7.674$ Å~\cite{schellenberg2010121sb}. $\mathcal{P}$ symmetry is present in this structure. The Eu, Mn, and Sb atoms occupy the Wyckoff positions 1a, 2d, and 2d, respectively. The crystallographic framework arising from this atomic arrangement is presented in Fig.~1. Fig. 1(a) presents the side view of the conventional unit cell, revealing the layered arrangement of Mn–Sb polyhedral networks separated by Eu layers. The top view, shown in Fig. 1(b), highlights the trigonal arrangement of Eu atoms and the hexagonal coordination within the Mn–Sb framework. Figure 1(c) illustrates the three-dimensional (3D) first Brillouin zone (BZ) with the corresponding high-symmetry path, along with the projected two-dimensional BZs for the (001) and (010) planes.

In the EuMnZnSb$_2$ system, one Mn atom in EuMn$_2$Sb$_2$ is substituted by Zn to investigate the effect of chemical doping on the structural and electronic properties. The symmetry of the relaxed structural and its Structural stability were analyzed using VASP, with post-processing performed through VASPKIT~\cite{wang2021vaspkit} and PHONOPY~\cite{togo2015first}. The optimized EuMnZnSb$_2$ compound crystallizes in a trigonal structure with space group P3m1 (No. 156), belonging to the hP Bravais lattice and point group $C{3v}$. Due to the Mn–Zn substitution, $\mathcal{P}$ symmetry is broken in this system.

The lattice parameters were initially adopted from the parent EuMn$_2$Sb$_2$ compound and subsequently optimized. The relaxed lattice constants are $a = b = 4.527$ Å and $c = 7.571$ Å, corresponding to a unit-cell volume of $V = 134.36$ Å$^{3}$. The equilibrium volume is comparable to that of pristine EuMn$_2$Sb$_2$, indicating that the substituted compound retains structural stability upon Zn incorporation.

To gain insight into the thermodynamically stable EuMnZnSb$_2$ system, we calculate the formation energy using the relation $E_f = E_{total}-E_{Eu}-E_{Mn}-E_{Zn}-2E_{Sb}$, which yields a value of -1.99 eV. Here, $E_{total}$ represents the total energy of EuMnZnSb$_2$ system, while $E_{Eu}, E_{Mn}, E_{Zn},$ and $E_{Sb}$ correspond to the total energies of isolated Eu, Mn, Zn, and Sb single atom, respectively. The negative formation energy indicates that EuMnZnSb$_2$ is thermodynamically stable. Furthermore, the dynamic stability was examined through phonon spectrum calculations. As shows in Fig.~2, all the phonon frequencies are positive throughout the entire Brillouin zone, confirming the dynamic stability of the EuMnZnSb$_2$ system. The projected phonon density of states (PDOS) reveals that the low-frequency region are mainly contributed by Eu atoms, whereas Mn, Zn, and Sb dominate the intermediate and high-frequency regions, consistent with their atomic masses.

\begin{figure*}
    \centering
    \includegraphics[width=1\linewidth]{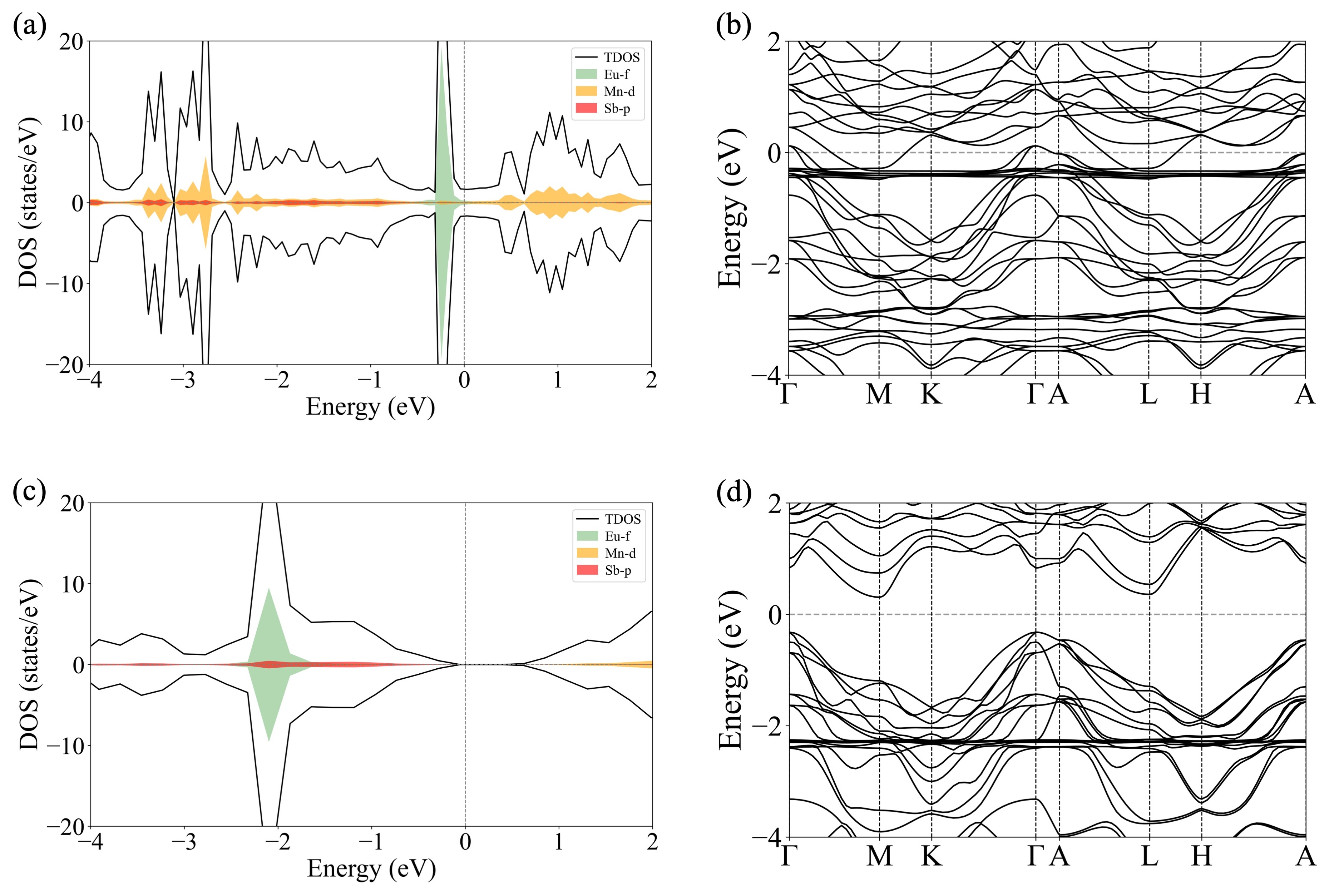}
    \caption{Electronic structure of EuMn$_2$Sb$_2$. Spin-polarized density of states (DOS) and band structures calculated within GGA and GGA+U. (a,b) DOS and band structure for the G-type antiferromagnetic ground state within GGA. (c,d) Corresponding results for the C-type antiferromagnetic state within GGA+U. The Fermi level is set to zero energy.}
    \label{fig:placeholder}
\end{figure*}

\begin{figure*}
    \centering
    \includegraphics[width=1\linewidth]{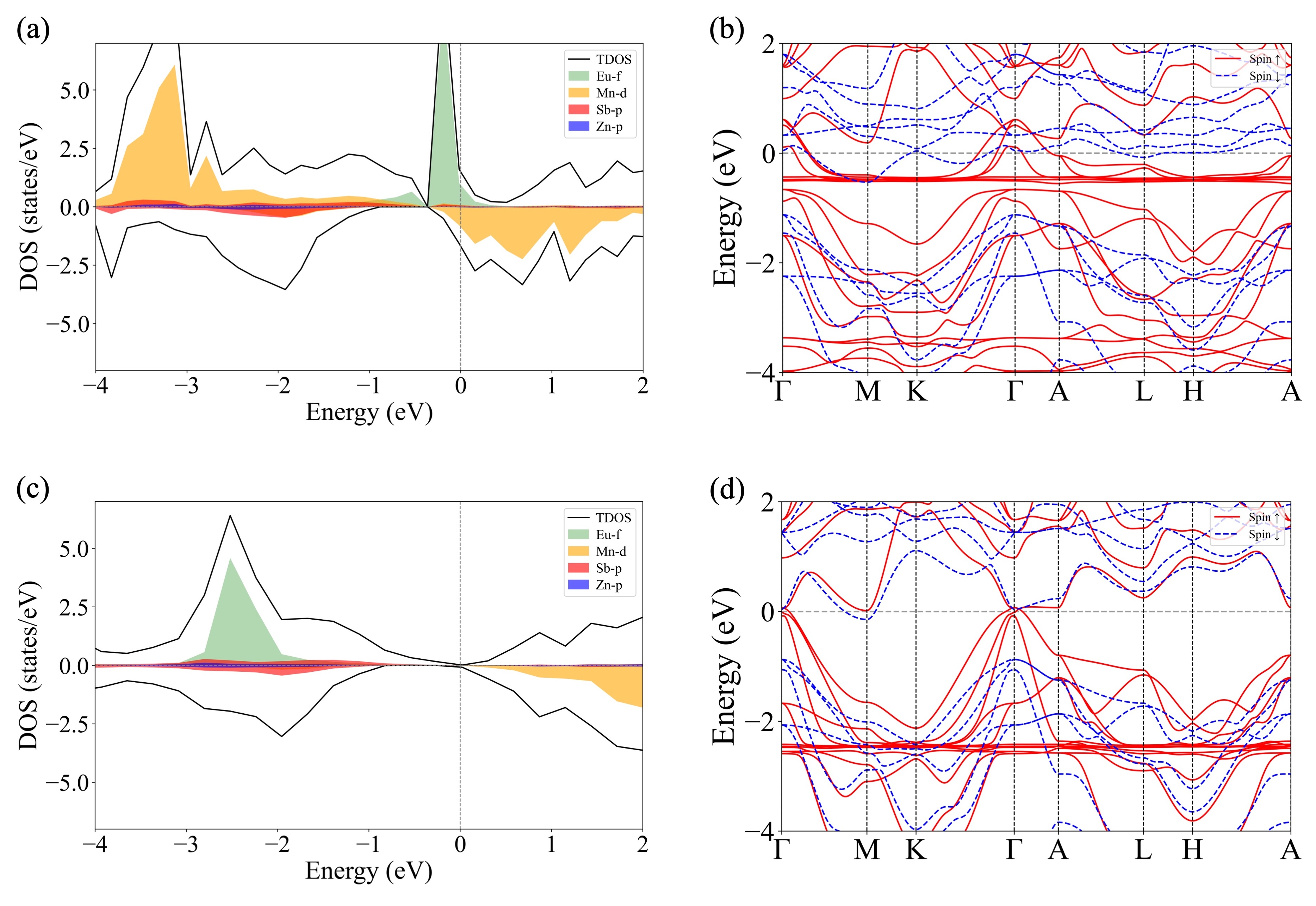}
    \caption{Electronic structure of EuMnZnSb$_2$ in the ferromagnetic state. Spin-polarized density of states (DOS) and band structures calculated within GGA and GGA+U. (a,b) DOS and band structure within GGA. (c,d) Corresponding results within GGA+U. The Fermi level is set to zero energy.}
    \label{fig:placeholder}
\end{figure*}
\subsection{\label{sec:citeref}Electronic Structure}
In this section, we discuss the electronic structure of two kinds of crystal structures, EuMn$_2$Sb$_2$ and Zn-substituted EuMnZnSb$_2$ compounds within GGA and GGA+$U$ frameworks. The analysis includes their spin-polarized density of state (DOS) and band structure. Fig.~3(a) shows the total and projected density of state (PDOS) of EuMn$_2$Sb$_2$ calculated within GGA. The results show that the Eu-$4f^7$ orbitals strongly dominated the total DOS near the Fermi level $(E_F)$ from -0.45 eV to 0.075 eV, and the Mn-$3d^5$ orbitals are mainly localized around the $E_F$ and are predominantly distributed within the valence band (VBs) between -3.5 eV and -0.5 eV. In contrast, the Sb-$6p$ orbitals are predominantly distributed in the occupied state in the energy window from -3.5 eV to -0.6 eV. To gain a clearer understanding of electronic structure, we calculated the band structure of EuMn$_2$Sb$_2$ calculated within the GGA frameworks, as shown in Fig.~3(b). The results indicate that the Eu-$4f^7$ and Mn-$3d^5$ orbitals are strongly localized near the $E_F$. Notably, three conduction bands  (CBs) intersect the $E_F$ along the high-symmetry directions. The bands near $E_F$ exhibit relatively weak dispersion in certain regions, indicating partial localization of electronic states, which is consistent with the dominant contribution of Eu-$4f^7$ and Mn-$3d^5$ orbitals as observed in DOS. These band crossings confirm that EuMn$_2$Sb$_2$ system exhibit metallic behavior within GGA framework. 

Since EuMn$_2$Sb$_2$ contains strongly correlated orbitals, particularly half field Eu-$4f^7$ and partially field Mn-$3d^5$ states, we further investigated its electronic structure within the GGA+$U$ framework. The on-site Coulomb interaction was treated using the Hubbard $U$ approach, with $U_{Eu}= 6.30$ eV and $U_{Mn}=5.50$ eV. These Hubbard parameters were determined using linear responses (LR) method~\cite{cococcioni2005linear} to ensure a reliable description of the localized electronic states. Figures 3(c) and 3(d) present the total and PDOS and the corresponding band structure of EuMn$_2$Sb$_2$ calculated within GGA+$U$, respectively. Compared to the GGA results, the inclusion of Hubbard $U$ significantly modifies the electronic structure. The Eu-$4f^7$ orbitals shift away from the $(E_F)$ and are predominantly distributed within the VBs from -2.5~eV to -1.5 eV,indicating enhanced localization. Meanwhile, the Mn-$3d^5$ and Sb-$6p$ orbitals are de-localized around the $E_F$, as shown in Fig.~3(c). The band structure shown in Fig.~3(b) reveals that no bands cross the $E_F$ along the high-symmetry directions. An indirect band gap of approximately 0.628 eV is observed, demonstrating that EuMn$_2$Sb$_2$ exhibits semiconducting behavior within GGA+$U$ framework.

We next discuss the electronic structure of the Zn-substituted compound EuMnZnSb$_2$ within the GGA and GGA+$U$ frameworks. The spin-polarized total and projected DOS calculated within GGA are shown in Fig.~4(a). The results indicate that the Eu-$4f^7$ and Mn-$3d^5$ orbitals predominant contribute to states near the Fermi level ($E_F$), while the Eu-$4f^7$ states are mainly localized slightly below $E_F$. The Sb-$6p$ states are primarily distributed within the VBs and exhibit a noticeable hybridization with the Mn-$3d^5$ states. In addition, the Zn states are mainly located in the deeper valence region and contribute weakly near the $E_F$, suggesting that the substitution of Zn modifies the electronic structure primarily through hybridization effects rather than direct contributions at $E_F$.

To further elucidate the band dispersion, the spin-polarized band structure calculated within GGA is shown in Fig.~4(b). Several bands are observed to cross the $E_F$ along the high-symmetry path, confirming the metallic character of EuMnZnSb$_2$ within the GGA approximation. The bands near $E_F$ exhibit moderate dispersion, indicating the itinerant nature of Mn-$3d^5$ electrons with partial contributions from Sb-$6p$ states. 

Taking into account the presence of strongly correlated Eu-$4f^7$ and Mn-$3d^5$ orbitals, we further examined the electronic structure within the GGA+$U$ framework. We used the Hubbard $U$ parameters same as EuMn$_2$Sb$_2$, obtained from the LR method, namely $U_{\mathrm{Eu}} = 6.30$ eV and $U_{\mathrm{Mn}} = 5.50$ eV. The corresponding DOS and band structures are presented in Figs.~4(c) and 4(d), respectively. Figures 4(c) and 4(d) present the total and PDOS and the corresponding band structure of EuMnZnSb$_2$ calculated within GGA+$U$, respectively. Upon inclusion of the Hubbard $U$, the Eu-$4f^7$ states shift toward lower energies and become more localized within the VBs from -3.1~eV to -1.5 eV, indicating enhanced correlation effects. The Mn-$3d^5$ states are also modified, leading to a noticeable reconstruction of the CBs near the $E_F$. Meanwhile, the Zn-$3p$ and Sb-$6p$ orbitals are mainly localized around the $E_F$, as shown in Fig 4(c). 
The calculated spin-polarized band structure shown in Fig.~4(b) reveals distinct electronic behaviors for the two spin channels. In the majority spin channel, no $E_g$ is observed, and the VBs and CBs touch near the $E_F$ along the $\Gamma$--A  high-symmetry directions, indicating metallic behavior. In contrast, the minority spin channel exhibits semiconducting character with an indirect band gap of approximately $0.7276$~eV between the VBs and CBs. The coexistence of metallic behavior in one spin channel and a finite band gap in the other indicates that the EuMnZnSb$_2$ exhibits half-metallic character, which is desirable for spintronic applications.

\begin{table}[h!]
\centering
\setlength{\tabcolsep}{8pt}
\caption{Calculated Hubbard parameters $U_{\mathrm{eff}}$ (eV), total energies (eV) of EuMn$_2$Sb$_2$ and EuMnZnSb$_2$ for FM and different AFM configurations within GGA and GGA+$U$, and total magnetic moments ($\mu_B$/atom).}
\label{tab:EuMnSb2}
\begin{tabular}{lcccc}
\toprule
 & \multicolumn{2}{c}{EuMn$_2$Sb$_2$} & \multicolumn{2}{c}{EuMnZnSb$_2$} \\
\cmidrule(lr){2-3}\cmidrule(lr){4-5}
 & GGA & GGA+$U$ & GGA & GGA+$U$ \\
\midrule
$U_{\mathrm{eff}}$ (Eu) & & 6.30 & & 6.30 \\
$U_{\mathrm{eff}}$ (Mn) & & 5.50 & & 5.50 \\
\midrule
FM                  & $-75.33$ & $-67.66$ & $-61.44$ & $-57.14$ \\
A-AFM               & $-75.40$ & $-67.68$ & $-61.42$ & $-57.12$ \\
C-AFM               & $-76.46$ & $-67.94$ & --- & --- \\
G-AFM               & $-76.47$ & $-67.93$ & --- & --- \\
$\Delta E$          & $-1.14$ & $-0.28$ & $0.02$ & $0.02$ \\
\midrule
$E_g$               & 0.00  & 0.628 & 0.00   & 0.00 \\
$\mu_{\mathrm{Eu}}$ & 6.767 & 6.973 & 6.859  & 7.006 \\
$\mu_{\mathrm{Mn}}$ & 3.829 & 4.620 & 3.981  & 4.581 \\
$\mu_{\mathrm{total}}$ & 0.00 & 0.00 & 2.163  & 2.304 \\
\bottomrule
\end{tabular}
\end{table}

\subsection{\label{sec:citeref}Magnetic Properties}
Conventionally, Mn-based layered pnictides exhibit thermoelectric and topological properties. Magnetic properties are evaluated by comparing the total energies obtained from spin-polarized ferromagnetic and antiferromagnetic calculations, along with analyzing the magnetic moments of the systems~\cite{Shakil2021}. In ferromagnetic materials, the absolute magnetization is identical to the total magnetization. In contrast, for antiferromagnetic systems, the total magnetization is zero, while the absolute magnetization corresponds to twice the sum of the magnetic moments of each pair of atoms in the system. The total energies of four magnetic structures with ferromagnetic (FM) and three antiferromagnetic (AFM) structure, namely A-type antiferromagnetic (A-AFM), C-type antiferromagnetic (C-AFM), and G-type antiferromagnetic (G-AFM) are summarized in table I. The magnetic configuration models used in this study were adopted from our previous work~\cite{sagar2026tunable}. 

Table~1 summarizes the total energies of all considered magnetic configurations, the magnetic energy differences, the total magnetic moments, and the local magnetic moments per Eu and Mn atom. The magnetic energy difference, $\Delta E$, represents the total exchange interaction and is defined as the energy difference between the AFM and FM configurations, $\Delta E = E_{\mathrm{AFM}} - E_{\mathrm{FM}}$. The calculated results indicate that the FM state is unstable for the EuMn$_2$Sb$_2$ system within both the GGA and GGA+$U$ approaches. Within GGA, EuMn$_2$Sb$_2$ stabilizes in the G-AFM configuration with a zero total magnetic moment. A significant negative $\Delta E$ of $-1.14$~eV is obtained, indicating strong AFM coupling and robust magnetic interactions in the system. The calculated local magnetic moments are $6.767~\mu_{\mathrm{B}}$ for Eu ($4f^7$) and $3.829~\mu_{\mathrm{B}}$ for Mn ($3d^5$). When on-site Coulomb interactions are included (GGA+$U$), the ground state changes to the C-AFM configuration, which also exhibits a zero total magnetic moment. The corresponding negative $\Delta E$ is $-0.28$~eV, still indicating strong AFM coupling. The calculated local magnetic moments increase to $6.973~\mu_{\mathrm{B}}$ for Eu and $4.620~\mu_{\mathrm{B}}$ for Mn. These values show good agreement with the reported experimental results. Therefore, the EuMn$_2$Sb$_2$ system exhibits an AFM ground state at room temperature ($T = 300$~K).

We next discuss the magnetic properties of the Zn-substituted compound EuMnZnSb$_2$ within the GGA and GGA+$U$ frameworks. The calculated results indicate that the FM state is energetically stable for the EuMnZnSb$_2$ system within both the GGA and GGA+$U$ approaches. Within GGA, EuMnZnSb$_2$ stabilizes in the FM configuration with a total magnetic moment of $2.163~\mu_{\mathrm{B}}$. A positive $\Delta E$ of $0.02$~eV is obtained, indicating ferromagnetic coupling and moderate magnetic interactions in the system. The calculated local magnetic moments are $6.859~\mu_{\mathrm{B}}$ for Eu ($4f^7$) and $3.981~\mu_{\mathrm{B}}$ for Mn ($3d^5$). When GGA+$U$, the FM configuration remains the ground state. The total magnetic moment remains $2.163~\mu_{\mathrm{B}}$, and the calculated $\Delta E$ is nearly identical to that obtained within GGA. The local magnetic moments increase to $7.006~\mu_{\mathrm{B}}$ for Eu and $4.581~\mu_{\mathrm{B}}$ for Mn. These results suggest that the EuMnZnSb$_2$ system exhibits a stable ferromagnetic ground state.

In the EuMn$_2$Sb$_2$ system, the Mn and Sb atoms form two-dimensional [Mn$_2$Sb$_2$]$^{\delta-}$ networks, which are charge-balanced and separated by layers of Eu atoms~\cite{schellenberg2010121sb}. Within these layers, Mn atoms carry localized $3d$ magnetic moments and are interconnected through Sb atoms, forming Mn--Sb--Mn exchange pathways. Hybridization between Mn-$3d$ and Sb-$p$ orbitals enables superexchange interactions that govern the magnetic ordering of the Mn sublattice. According to the Goodenough–Kanamori–Anderson rules~\cite{goodenough1955theory,kanamori1957theory,anderson1950antiferromagnetism}, such superexchange interactions often favor AFM coupling depending on the Mn--Sb--Mn bond geometry and orbital overlap. In contrast, the Eu atoms mainly act as spacer layers between the [Mn$_2$Sb$_2$]$^{\delta-}$ networks and host localized Eu-$4f^7$ moments, which interact more weakly with the Mn sublattice due to the highly localized nature of the $4f$ orbitals. As a result, the magnetic behavior of EuMn$_2$Sb$_2$ is primarily governed by Mn--Sb--Mn exchange interactions within the two-dimensional layers, leading to stabilization of the AFM ground state.

The substitution of Zn in EuMn$_2$Sb$_2$ significantly modifies the magnetic ground state by altering the electronic structure and magnetic exchange interactions within the two-dimensional [Mn$_2$Sb$_2$]$^{\delta-}$ networks. In the parent compound EuMn$_2$Sb$_2$, the superexchange Mn--Sb--Mn interactions favor AFM coupling within the layered [Mn$_2$Sb$_2$]$^{\delta-}$ networks. Upon Zn substitution, the chemical environment of the [Mn$_2$Sb$_2$]$^{\delta-}$ layers is modified, which alters the Mn--Sb--Mn exchange pathways and the hybridization between Mn-$3d$ and Sb-$p$ states. This modification of the electronic structure near the $E_F$ weakens the AFM superexchange interaction and can enhance FM exchange contributions. Consequently, the delicate balance between competing magnetic interactions within the [Mn$_2$Sb$_2$]$^{\delta-}$ networks shifts, leading to the stabilization of a ferromagnetic ground state in the Zn-substituted system. Such chemical tuning of magnetic exchange interactions has been widely reported in Mn-based layered pnictides, where substitution can modify the electronic structure and exchange pathways, resulting in changes in magnetic ordering.

\begin{figure*}
    \centering
    \includegraphics[width=1\linewidth]{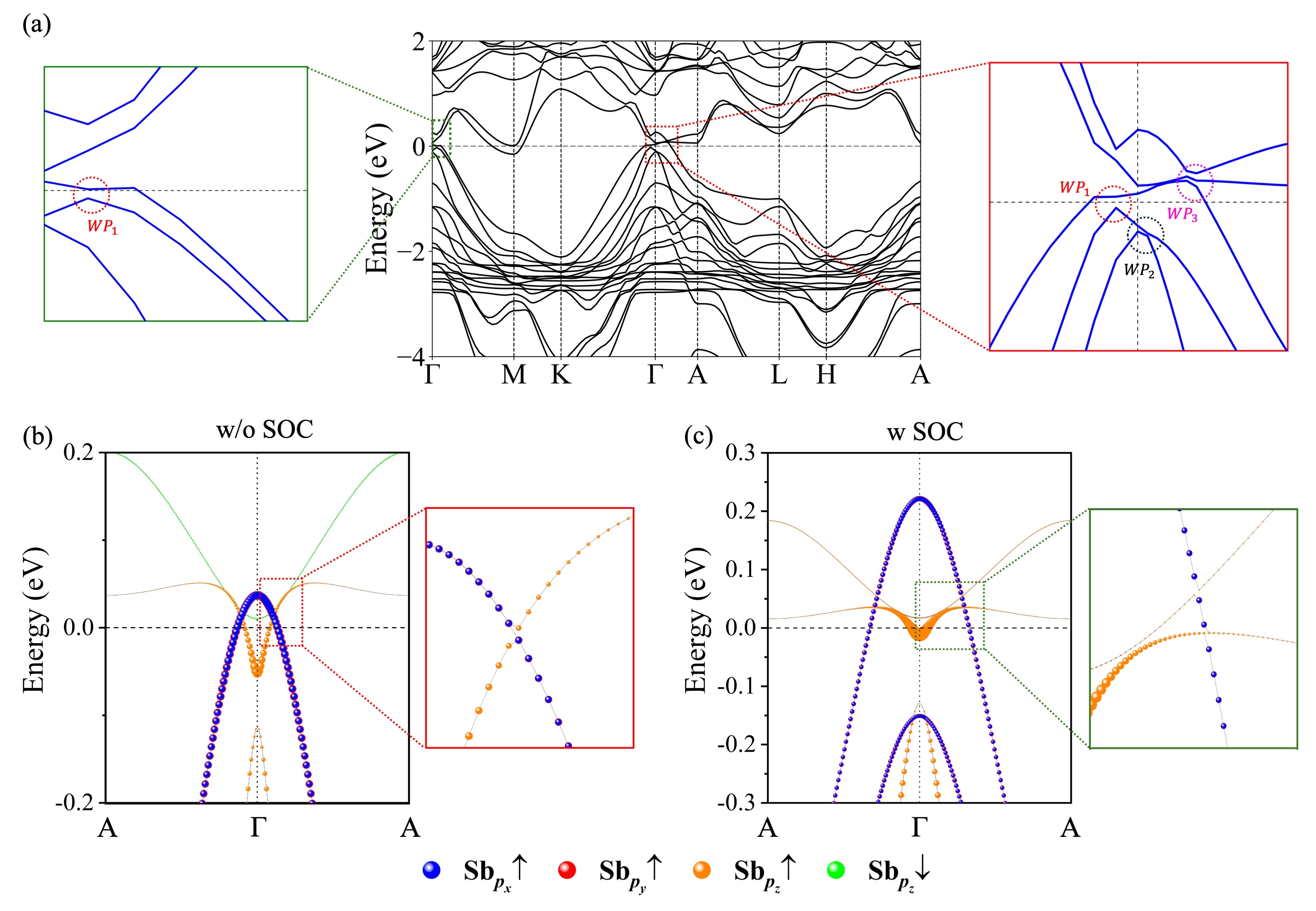}
    \caption{(a) Calculated electronic band structure including SOC. Several band crossings near the $E_F$, giving rise to Weyl nodes, where the Weyl nodes $WP_1$, $WP_2$, and $WP_3$ are indicated. Band inversion along the $A$--$\Gamma$--$A$ high-symmetry path without (b) and with (c) SOC.}
    \label{fig:placeholder}
\end{figure*}

\begin{figure*}
    \centering
    \includegraphics[width=1\linewidth]{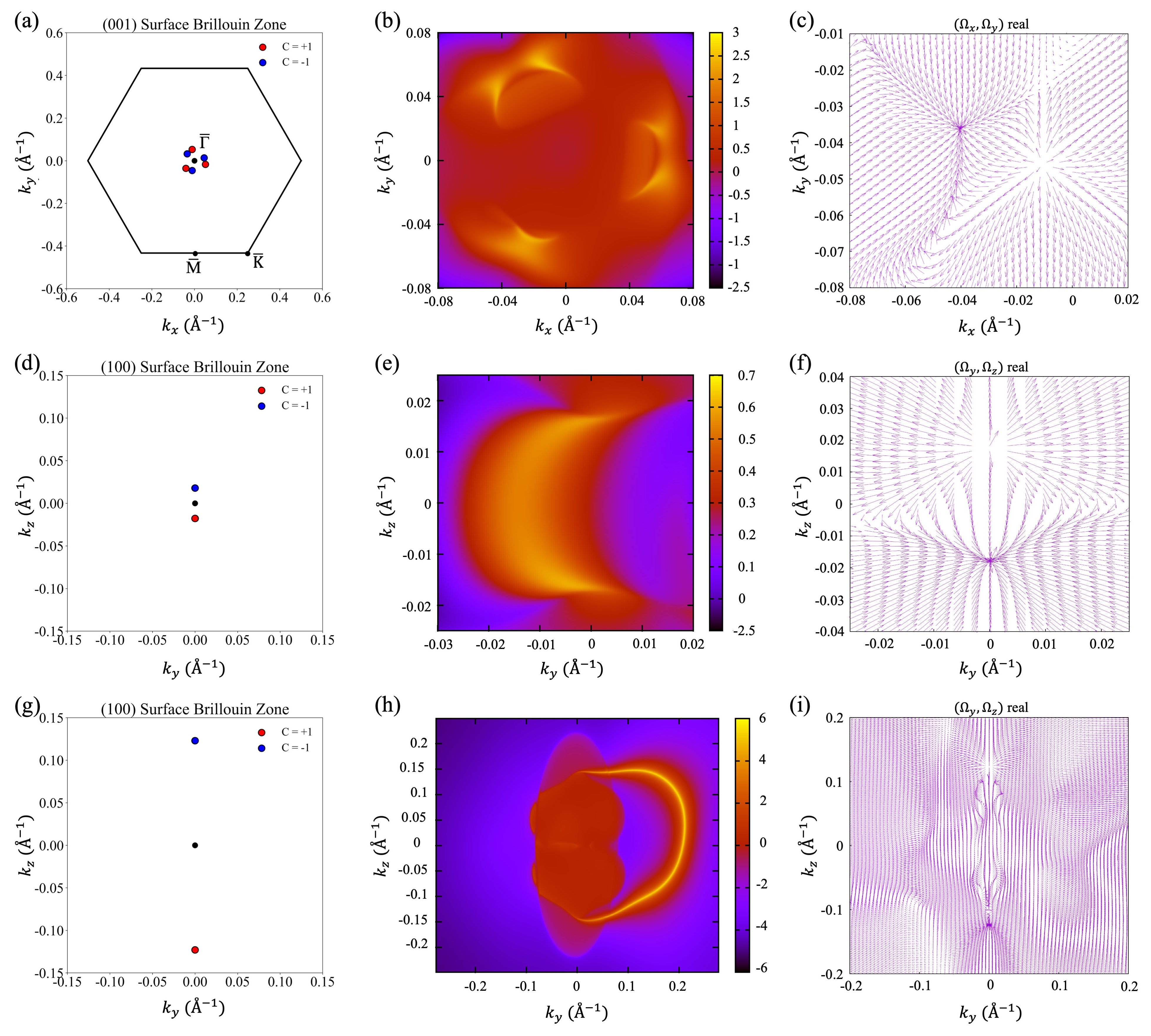}
    \caption{Positions of the Weyl nodes in the Brillouin zone (BZ), the associated Fermi arcs on the surface states, and the Berry curvature distribution in momentum space. (a)–(c) Weyl nodes pair $WP_1$: (a) projected positions of the Weyl nodes with opposite chiralities ($C=\pm1$) in the (001) surface BZ. (b) Surface spectral function showing the corresponding Fermi arc connecting the projections of the Weyl nodes. (c) Berry curvature distribution $(\Omega_x,\Omega_y)$ in momentum space, where the Weyl nodes act as monopole sources and sinks of Berry curvature. (d)–(f) Weyl nodes pair $WP_2$: (d) projected positions of the Weyl nodes in the (100) surface BZ. (e) Surface spectral function revealing the Fermi arc connecting the projections of the Weyl nodes. (f) Berry curvature vector field $(\Omega_y,\Omega_z)$ illustrating the characteristic source–sink structure associated with the Weyl nodes. (g)–(i) Weyl nodes pair $WP_3$: (g) projected locations of the Weyl nodes in the (100) surface BZ. (h) Surface spectral function showing the corresponding Fermi arc connecting the Weyl nodes with opposite chirality. (i) Berry curvature distribution $(\Omega_y,\Omega_z)$ confirming the monopole nature of the Weyl nodes, where Berry curvature flows outward from the node with $C=+1$ and inward toward the node with $C=-1$.}
    \label{fig:placeholder}
\end{figure*}

\subsection{\label{sec:citeref}Magnetic Weyl Semimetal of EuMnZnSb$_2$}

Weyl semimetals (WSMs) can generally be classified into two categories: magnetic WSMs and noncentrosymmetric WSMs, which arise from the breaking of either $\mathcal{T}$ or $\mathcal{P}$, respectively. In this section, we propose a different type of WSMS phase in which both $\mathcal{T}$ and $\mathcal{P}$ symmetries are simultaneously broken. Fig. 4(d) show the band structure within GGA+$U$. In the calculated band structure, the majority spin bands near the $E_F$ are mainly dominated by Sb-$p_{x,y}$ and Sb-$p_z$ orbitals. The VBs and CBs touch each other and form a Dirac-cone-like dispersion along the $\Gamma$–A high-symmetry direction in the BZ. Since the compound contains heavy elements such as Eu and Sb, characterized by Eu-$4f$ and Sb-$6p$ orbitals, the effect of spin–orbit coupling (SOC) is expected to be significant. Upon inclusion of SOC, the majority and minority spin channels become mixed, and the Eu-$4f$ and Sb-$6p$ bands undergo spin–orbit splitting as shown in Fig. 5(a). This splitting lifts the band degeneracies and modifies the electronic structure near the $E_F$.

To gain deeper insight into the orbital contributions to the band structure, we examine the evolution of the electronic states near the $E_F$ along the $A$--$\Gamma$--$A$ high-symmetry path, both without and with SOC. In the absence of SOC [Fig.~5(b)], the electronic states near $E_F$ are predominantly derived from the Sb-$p$ orbitals. In particular, the Sb-$p_{x,y}$ orbitals exhibit significant overlap close to the $E_F$. The hybridization between the Sb-$p_{x,y}$ and Sb-$p_{z}$ orbitals gives rise to a Dirac-cone-like linear band dispersion near $E_F$, as illustrated in the inset of Fig.~5(b), indicating symmetry-protected band crossings along the high-symmetry direction. Upon inclusion of SOC, the electronic structure undergoes a noticeable modification. The relative ordering of the Sb-$p_{x,y}$ and Sb-$p_{z}$ states changes significantly due to the strong relativistic interaction. As a consequence, the bands originating from Sb-$p_{x,y}$ and Sb-$p_{z}$ orbitals intersect near the $E_F$ and exhibit a pronounced band inversion, as shown in Fig.~5(c) and its inset. This SOC-induced rearrangement of orbital characters highlights the important role of SOC in shaping the low-energy electronic structure of the system. Such SOC-driven band inversion is often regarded as a hallmark of nontrivial topological electronic states in materials containing heavy elements such as Sb.

\begin{table}[h]
\centering
\setlength{\tabcolsep}{4pt}
\caption{Momentum-space coordinates ($k_x$, $k_y$, $k_z$), energies relative to the Fermi level (in eV), and corresponding chiralities of the Weyl nodes.}
\begin{tabular}{cccccc}
\toprule
WP & $k_x$ & $k_y$ & $k_z$ & Energy (eV) & Chirality \\
\midrule
$WP_1$ & -0.0403 & -0.0358 & -0.0006 &  0.0015  &  1 \\
       & -0.0110 & -0.0459 & -0.0029 & -0.0110  & -1 \\

$WP_2$ &  0      &  0      &  0.1229 &  0.0785  & -1 \\
       &  0      &  0      & -0.1229 &  0.0785  &  1 \\

$WP_3$ &  0      &  0      & -0.0177 & -0.1120  &  1 \\
       &  0      &  0      &  0.0178 & -0.1120  & -1 \\
\bottomrule
\end{tabular}
\label{tab:weyl_points}
\end{table}

Since SOC is included, the breaking of $\mathcal{T}$ caused by ferromagnetism and the lack of $\mathcal{P}$ lift band degeneracies and generate pairs of Weyl nodes with opposite chirality in momentum space. Fig.~5(a) shows the calculated band structure including SOC. Upon inclusion of SOC, several bands crossing emerge near the $E_F$ that give rise to Weyl nodes. In particular, we identify a pair of Weyl nodes formed by the crossing between the VBs and CBs close to the $E_F$. These Weyl nodes possess opposite chirality $C=\pm1$, as required by the topological charge conservation in momentum space. One of these nodes is located slightly above the $E_F$ at $0.0015$~eV, while the other appears below the $E_F$ at $-0.0110$~eV. 

Furthermore, additional Weyl nodes are found from band crossings within the same bands manifold. Specifically, one pair of Weyl nodes emerges from the crossing of two VBs, and another pair arises from the crossing of two CBs. The corresponding energies relative to the $E_F$ are $-0.1120$~eV and $0.0758$~eV, respectively. All these Weyl nodes occur in pairs and carry opposite topological charge (chirality) characterized by a Chern number $C=\pm1$ in momentum space, consistent with the Nielsen--Ninomiya theorem~\cite{nielsen1981absence,zhao2016novel}. The locations of these Weyl nodes and their corresponding band crossings are highlighted in Fig.~5(a) and its insets. The coordinates, energies, and corresponding chiralities of these Weyl nodes are presented in Table~II.

Figure~6 illustrates the momentum-space positions of the Weyl nodes, the corresponding surface Fermi arcs, and the Berry curvature distribution associated with each WP. Figures~6(a)--6(c) correspond to the Weyl pair $WP_1$. Fig.~6(a) shows the projected positions of the Weyl nodes in the (001) surface BZ, where the nodes with opposite chiralities ($C=\pm1$) are located close to the $\bar{\Gamma}$ point. The corresponding surface spectral function shown in Fig.~6(b) reveals the presence of surface Fermi arcs connecting the projections of the Weyl nodes. The Berry curvature distribution in Fig.~6(c) exhibits a clear monopole-like pattern in momentum space, where the Weyl nodes act as sources and sinks of Berry curvature, reflecting their opposite topological charges.

Figures~6(d)--6(f) correspond to the Weyl pair $WP_2$. Fig.~6(d) shows the projected positions of the Weyl nodes in the (100) surface BZ. The associated surface spectral intensity in Fig.~6(e) demonstrates a Fermi arc connecting the projections of the two Weyl nodes. The Berry curvature vector field shown in Fig.~6(f) again displays a characteristic source--sink structure centered at the Weyl nodes, confirming their nontrivial topological nature.

Figures~6(g)--6(i) correspond to the Weyl pair $WP_3$. Fig.~6(g) presents the locations of the Weyl nodes projected onto the (100) surface BZ. The surface spectral function in Fig.~6(h) shows the corresponding Fermi arc connecting the two Weyl nodes with opposite chirality. The Berry curvature distribution in Fig.~6(i) further confirms the monopole behavior of the Weyl nodes, where the WP with $C=-1$ acts as a source of Berry curvature, while the WP with $C=+1$ behaves as a sink.

The calculated electronic structure reveals that EuMnZnSb$_2$ hosts multiple pairs of Weyl nodes located close to the $E_F$, indicating that this compound realizes a MWSM phase. In this system, the simultaneous breaking of $\mathcal{P}$ symmetry, caused by Zn substitution, and $\mathcal{T}$ symmetry, arising from ferromagnetic ordering in the presence of SOC, lifts band degeneracies and stabilizes isolated band crossing in momentum space. Such symmetry conditions are well known to generate Weyl nodes in topological semimetals and have been widely discussed in both theoretical and experimental studies of WSMs~\cite{armitage2018weyl,zou2019study}.

Our calculations identify three distinct pairs of Weyl nodes ($WP_1$, $WP_2$, and $WP_3$) distributed in the BZ near the $E_F$. These nodes possess opposite chiralities ($C=\pm1$), consistent with the Nielsen–Ninomiya theorem, which requires that Weyl nodes in a periodic lattice must occur in pairs with opposite topological charge so that the total chirality in the BZ vanishes~\cite{nielsen1981absence,zhao2016novel}. The presence of Weyl nodes close to the $E_F$ is particularly important because it allows the topological quasiparticles to strongly influence the low-energy electronic and transport properties of the material.

The topological nature of the Weyl nodes is confirmed by the calculated Berry curvature distribution, which exhibits monopole-like structures in momentum space where each WP acts as a source or sink of Berry curvature depending on its chirality. These Berry curvature monopoles represent the fundamental topological invariant of Weyl fermions and strongly influence transport responses in WSMs, leading to phenomena such as anomalous Hall effects and nonlinear optical responses~\cite{armitage2018weyl,pandey2024ab,burkov2016topological}. In addition, our surface spectral calculations reveal open Fermi arc surface states connecting the projections of Weyl nodes with opposite chirality on the surface BZ. Such Fermi arcs originate from the bulk–boundary correspondence of topological band structures and have been experimentally observed in several WSMs including the TaAs family and MWSMs~\cite{xu2015discovery,wang2016time,belopolski2019discovery}. The coexistence of Berry curvature and Fermi arc surface states therefore provides strong evidence for the nontrivial topology of EuMnZnSb$_2$.

MWSMs are particularly interesting because broken $\mathcal{T}$ symmetry can generate strong Berry curvature near the Weyl nodes, leading to large anomalous Hall conductivity~\cite{armitage2018weyl,zou2019study}. For example, the MWSM Co$_3$Sn$_2$S$_2$ exhibits a giant anomalous Hall effect arising from Berry curvature concentrated around its Weyl nodes~\cite{lohani2023electronic}. Since several Weyl nodes in EuMnZnSb$_2$ lie close to the $E_F$, similar Berry-curvature-driven transport responses may occur in this system. Moreover, Weyl fermions can produce the chiral anomaly under parallel electric and magnetic fields, which experimentally manifests as negative longitudinal magnetoresistance~\cite{armitage2018weyl}. Compared with previously reported WSMs, nonmagnetic systems such as the TaAs family host Weyl nodes mainly due to $\mathcal{P}$ symmetry breaking combined with strong SOC~\cite{xu2015discovery}, whereas MWSMs such as Co$_3$Sn$_2$S$_2$ and the RAlGe family host Weyl nodes generated by magnetic ordering that breaks $\mathcal{T}$ symmetry~\cite{chang2018magnetic}. Similar to these systems, EuMnZnSb$_2$ hosts Weyl nodes originating from the combined effects of broken symmetries and magnetism.


\section{Conclusion}

In this work, we systematically investigated the structural, electronic, magnetic, and topological properties of EuMn$_2$Sb$_2$ and the Zn-substituted compound EuMnZnSb$_2$ using first-principles density functional theory calculations. Our results reveal that pristine EuMn$_2$Sb$_2$ exhibits a semiconducting ground state with an energy gap of approximately 0.628 eV and stabilizes in a C-type AFM configuration. The magnetic behavior is primarily governed by Mn--Sb--Mn superexchange interactions within the layered crystal structure.

Upon Zn substitution at the Mn site, the magnetic exchange interactions are significantly modified, resulting in a transition from AFM to FM ordering. In addition, Zn substitution breaks $\mathcal{P}$ symmetry in the crystal structure. When SOC is included, the simultaneous breaking of $\mathcal{P}$ and $\mathcal{T}$ symmetry leads to the emergence of a MWSM phase in EuMnZnSb$_2$. In this phase, multiple pairs of Weyl nodes with opposite chirality are identified in the vicinity of the $E_F$.

The nontrivial topological nature of these Weyl nodes is further confirmed through Berry curvature calculations, which reveal characteristic monopole-like source and sink distributions in momentum space. Moreover, topologically protected Fermi-arc surface states connecting the projected Weyl nodes are observed in the surface BZ, providing strong evidence for the WSM character of the system.

Overall, our findings demonstrate that chemical substitution is an effective strategy for tuning both magnetic ordering and electronic topology in EuMnZnSb$_2$, a compound belonging to the CaAl$_2$S$_2$-type pnictide family. The presence of Weyl nodess close to the $E_F$ suggests that EuMnZnSb$_2$ may exhibit intriguing transport phenomena, such as the anomalous Hall effect and chiral anomaly-induced magnetoresistance. These results highlight EuMnZnSb$_2$ as a promising platform for exploring magnetic Weyl physics and its potential applications in topological electronics and spintronic devices.

\bibliography{MX}

\end{document}